\documentclass[a4paper,12pt]{article}

\usepackage{graphics}

\begin{document}

\vspace{1cm}
\thispagestyle{empty}
\title{
\vspace{-2cm}
\begin{flushright}
{\normalsize IRB-TH-2/00 \\ 
\vspace{-0.5cm}
April 2000}
\end{flushright}
\vspace{1 cm}
\bf Lifetime-difference pattern of heavy hadrons}

\vspace{1cm}
\author{B. Guberina\thanks{E-mail: guberina@thphys.irb.hr},
 B. Meli\'{c}\thanks{E-mail: melic@thphys.irb.hr},
 H. \v Stefan\v ci\'c\thanks{E-mail: shrvoje@thphys.irb.hr}
}

\vspace{2cm}
\date{
\centering
Theoretical Physics Division, Rudjer Bo\v{s}kovi\'{c} Institute, \\
   P.O.Box 180, HR-10002 Zagreb, Croatia}


\maketitle

{\abstract The preasymptotic effects originating from the
four-quark operators are disentangled from the other contributions
by considering appropriate combinations of inclusive decay
rates. Under the assumption of the isospin and heavy-quark
symmetry, a set of
relations connecting charmed and beauty decays is obtained without
invoking specific models. The results are compared with other
approaches and confronted with experiment. }

\vspace{1cm}


\vspace{1cm}

It is astonishing that a lot of physical observables, decay rates
of semileptonic, nonleptonic, and radiative decays are described in
the framework of the inverse heavy-quark mass expansion in terms of
relatively few basic quantities, e.g., quark masses and hadronic
expectation values ($HEV$) of several leading local operators 
\cite{Bigi00,Neu,BlokShif}. The underlying theory is based on a 
few field-theory 
basics, such as the operator product expansion, quark-hadron
duality,
and certain well-known symmetries. Even more surprisingly, the
theory seems to work rather well in the case of charmed hadrons,
where the expansion parameter  
$\sqrt{\mu_{G}^{2}(D)/m_{c}^{2}} \simeq 0.5$ is by no means very
small. A systematic analysis leads to very clear 
lifetime-pattern 
predictions, the agreement with experiment being reasonable
even for absolute lifetime values \cite{GM,GMScc}.

In beauty hadron decays, one expects that the whole theory should
work much better, since the expansion parameter, 
$\sqrt{\mu_{G}^{2}(B)/m_{b}^{2}} \simeq 0.13 \ll 1$, 
is significantly smaller than in the case of charmed hadrons.
Still, in spite of the overall agreement between theory and
experiment \cite{Bigi00}, there are some questions to be answered.
The main question is: does the quark-hadron duality work? 
Unfortunately, for a reliable study of duality, one needs complete
control over nonperturbative phenomena, which is out of reach of 
the present theory. Nevertheless, the validity of duality was studied
in exactly solvable 't Hooft model in $QCD_{2}$ \cite{dual1,dual2}, 
leading to a perfect matching, which is certainly encouraging.
Still, one would be happy to find stronger support for duality in
the $(3+1)$ theory, too.

The question is: could one try to reduce the uncertanties of the
calculation, like the strong sensibility to the value of $m_{Q}$,
and/or find the way to disentangle various preasymptotic
effects which cause lifetime differences?

The decay rate of a decaying hadron is generally of the type 
\begin{equation}
\label{eq:master}
 \Gamma (H_{Q} \rightarrow f) = \frac{G_{\rm F}^2 m_Q^5}{192 \pi^3} |V|^2
\frac{1}{2 M_{H_{Q}}} \left[ \sum_{D=3}^{D_{c}} c_{D}^{f} 
\frac{\langle H_{Q} | O_{D} | H_{Q} \rangle}{m_{Q}^{D-3}}
+ {\cal O}(1/m_{Q}^{D_{c}-2}) \right] \, ,
\end{equation}
where $c_{D}^{f}$ are the Wilson coefficients and
$\langle H_{Q} | O_{D} | H_{Q} \rangle$ are the matrix elements of 
the $D$-dimensional operators which are suppressed by the inverse
power of mass $1/m_{Q}^{D-3}$.

The following comments are in order:
\begin{enumerate}
\item It would be desirable to extract a combination of rates
$\Gamma'$ that depends only on the operators with $D=6$, since they are
{\em the} operators responsible for lifetime differences.
\item Since the $D=6$ operators are suppressed by $m_{Q}^{-3}$
with respect to the leading $D=3$ operator, this would reduce the
mass dependence of $\Gamma'$ to $m_{Q}^{2}$, thus significantly reducing
the errors coming from the uncertainty of $m_{Q}$.
\item Even truncating the sum in (\ref{eq:master}) at $D=6$, one
would presumably be able to test the role of
preasymptotic effects with satisfactory accuracy,
relying on heavy-quark symmetry. That is the aim of this paper. 
\end{enumerate}  

 As both the decays in the 
$b$ and $c$ sectors of hadrons are described by the same formalism,
it seems worthwhile to investigate the possibility of establishing
connections between these two sectors. In a recent paper 
\cite{Voloshin}, Voloshin demonstrated one of such possible
connections. 
The strength of that approach is in
avoiding the model dependence assuming 
$SU(3)$ {\em flavor} and {\em heavy-quark symmetry (HQS)}. 
One can extend this approach to obtain more predictivity. Using
the matrix elements extracted from charmed baryons, a 
parameter $F_{B}^{eff}$ (parametrizing four-quark contributions to 
beauty baryons) can be determined and concrete numerical
predictions can be obtained in this way \cite{GMS}. This procedure
brings the ratio $\tau(\Lambda_{b})/\tau(B_{d}^{0})$ to better
agreement with experiment and gives predictions for the lifetimes
of beauty baryons. 

In the present paper we adopt a similar, yet different strategy from that
followed in \cite{Voloshin}. In \cite{Voloshin}, the author, when
establishing the connection between $c$ and $b$ baryons, always relates
baryons with the same light-quark content, i.e.,  
mutually related baryons
differ in the flavor of the heavy quark. Also, the author
explicitly extracts values of four-quark operator 
matrix elements and
then applies them elsewhere. Our approach differs in both these
respects. The heavy hadrons that we relate neither contain the
same heavy quark(s) nor the same light (anti)quarks -- they are 
affected by the same type of the four-quark operator contribution. 
Furthermore, we
form such combinations that the four-quark 
operator matrix elements get reduced.

In forming such combinations, we consider the leading modes of the decay of
$c$ and $b$ quarks with respect to the CKM matrix elements. 
For the $c$ quark, we then have only one nonleptonic mode, $c \rightarrow s
\overline{d}u$, and one semileptonic mode per lepton family,
$c \rightarrow s \overline{l} \nu_{l}$.
In the case of $b$ quark, there are two nonleptonic modes, 
$b \rightarrow c \overline{u} d$ and $b \rightarrow c \overline{c}
s$, as well as one semileptonic mode per lepton family,
$ b \rightarrow c l \overline{\nu_{l}}$. As we consider
$b$ baryons containing only light quarks along with $b$ quarks, the
semileptonic Cabibbo leading modes for the decay of the $b$ quark do not appear.
We also disregard mass corrections of four-quark operators coming
from the massive particles in final decay states. This is a better
approximation for $c$ decays (as $m_{s}^{2}/m_{c}^{2} \sim 0.01$)
than for $b$ decays (where $m_{c}^{2}/m_{b}^{2} \sim 0.1$).

Traditionally, the effects of four-quark operators are called
the positive Pauli interference, the negative Pauli interference
and, $W$ exchange or
annihilation \cite{cmesons,cbaryons}. 
We consider pairs of one charmed and one beauty particle
which contain the same type of four-quark contribution. 
It is important to notice that in $b$ and $c$ decays different
light-quark flavors participate in the same type of four-quark contributions.
Using {\em isospin ($SU(2)$ flavor)} symmetry and {\em HQS}
we are able to form appropriate decay-rate differences which lead us to the
final result. 

We start our considerations with heavy mesons. The first pair of
considered particles are $D^{+} \, (c\overline{d})$ and $B^{-} \,
(b\overline{u})$. In both these mesons the effect of negative
Pauli interference occurs. The contributions of 
four-quark operators to the
decay rates of these two particles are described by the same type of operators.
Moreover, the operators for the $b$ case can be obtained from those in
the $c$ case by making the substitution $c \rightarrow b$, $\overline{d} 
\rightarrow \overline{u}$. The second pair of particles form
$D^{0} \, (c\overline{u})$ and $B^{0} \, (b \overline{d})$. In both
these mesons the effect of $W$ exchange occurs. 

Next, we would like to isolate the effects of four-quark operators,
i.e., we have to eliminate the contributions of the operators of
dimensions $5$ and lower. We achieve this goal by taking differences
of the decay rates of $D$ and $B$ mesons assuming the {\em isospin}
symmetry. This procedure leads us to the following relations:
\begin{equation}
\label{eq:diffD}
\Gamma(D^{+}) - \Gamma(D^{0}) = \frac{G_{F}^{2} m_{c}^{2}}{4 \pi}
|V_{cs}|^{2} |V_{ud}|^{2} [ \langle D^{+} | P^{cd} | D^{+} 
\rangle - \langle D^{0} |  P^{cu} | D^{0} \rangle ]\, ,
\end{equation}
\begin{equation}
\label{eq:diffB}
\Gamma(B^{-}) - \Gamma(B^{0}) = \frac{G_{F}^{2} m_{b}^{2}}{4 \pi}
|V_{cb}|^{2} |V_{ud}|^{2} [ \langle B^{-} | P^{bu} | B^{-} 
\rangle - \langle B^{0} | P^{bd} | B^{0} \rangle ] \, ,
\end{equation}
where $P$'s denote the appropriate four-quark terms 
from the heavy-quark effective Lagrangian, see, e.g., Ref. \cite{Voloshin}.

Using {\em HQS} and {\em isospin} symmetry we
can express the terms
describing the negative Pauli interference as
\begin{equation}
\label{eq:HQSfornegintD}
P^{cd} = \sum_{i=1}^{2} c_{i}^{cd} (\mu) O_{i}^{cd} (\mu) =
\sum_{i=1}^{2} c_{i, negint}^{Qq} (\mu) O_{i, negint}^{Qq} (\mu)
+ {\cal O}(1/m_{c}) \, ,
\end{equation}
\begin{equation}
\label{eq:HQSfornegintB}
P^{bu} = \sum_{i=1}^{2} c_{i}^{bu} (\mu) O_{i}^{bu} (\mu) =
\sum_{i=1}^{2} c_{i, negint}^{Qq} (\mu) O_{i, negint}^{Qq} (\mu)
+ {\cal O}(1/m_{b}) \, ,
\end{equation}
where $Q$ denotes a heavy quark in the heavy-quark limit  
$m_{Q} \rightarrow \infty$. For the matrix elements of the above mentioned
operators we obtain (by applying {\em HQS} and {\em isospin}
symmetry again) $\langle D^{+} | P^{cd} | D^{+} \rangle = A_{negint} + 
{\cal O}'(1/m_{c})$ and $\langle B^{-} | P^{bu} | B^{-} \rangle 
= A_{negint} + {\cal O}'(1/m_{b})$,
%
%
%
where the matrix element is $A_{negint} = \langle M_{Q\overline{q}} |
\sum_{i=1}^{2} c_{i, negint}^{Qq} (\mu) O_{i, negint}^{Qq} (\mu)
| M_{Q\overline{q}} \rangle$ and $M_{Q\overline{q}}$ denotes the
mesonic state with one heavy quark $Q$ and one light antiquark 
$\overline{q}$ in the heavy-quark limit. In the preceding relations 
we have also stated that heavy-quark mass suppressed corrections to
the operators and their matrix elements do not
have to be the same.

In an analogous manner we can obtain the expressions for the operators 
describing $W$ exchange using the notation explained above:
\begin{equation}
\label{eq:HQSforexchD}
P^{cu} = \sum_{i=1}^{4} c_{i}^{cu} (\mu) O_{i}^{cu} (\mu) =
\sum_{i=1}^{4} c_{i, exch}^{Qq} (\mu) O_{i, exch}^{Qq} (\mu)
+ {\cal O}(1/m_{c}) \, ,
\end{equation}
\begin{equation}
\label{eq:HQSforexchB}
P^{bd} = \sum_{i=1}^{4} c_{i}^{bd} (\mu) O_{i}^{bd} (\mu) =
\sum_{i=1}^{4} c_{i, exch}^{Qq} (\mu) O_{i, exch}^{Qq} (\mu)
+ {\cal O}(1/m_{b}) \, .
\end{equation}
The matrix elements of the operators
given by (\ref{eq:HQSforexchD}) and (\ref{eq:HQSforexchB}) are
$\langle D^{0} | P^{cu} | D^{0} \rangle = A_{exch} + 
{\cal O}'(1/m_{c})$ and $\langle B^{0} | P^{bd} | B^{0} \rangle 
= A_{exch} + {\cal O}'(1/m_{b})$,
%
%
%
with $A_{exch} = \langle M_{Q\overline{q}} |
\sum_{i=1}^{4} c_{i, exch}^{Qq} (\mu) O_{i, exch}^{Qq} (\mu)
| M_{Q\overline{q}} \rangle$. In all relations given above, the
dependence of the coefficients $c_{i}$ and the operators $O_{i}$ on 
the scale
$\mu$ is explicitly displayed. 
Since the dependences of the operators
and the coefficients on $\mu$ cancel each other, there is no need
for an 
explicit value for the $\mu$ scale, and also no hybrid renomalization 
is required. That scale, however, should be small compared 
with the heavy-quark masses, to allow 
for the proper heavy-quark expansion and the reduction of the 
four-quark operator contribution in the heavy-quark limit, as 
shown below. 

Combining all preceding expressions, we obtain for the decay rate 
differences
\begin{equation}
\label{eq:finalD1}
\Gamma(D^{+}) - \Gamma(D^{0}) = \frac{G_{F}^{2} m_{c}^{2}}{4 \pi}
|V_{cs}|^{2} |V_{ud}|^{2} [A_{negint} - A_{exch} + {\cal
O}(1/m_{c})] \, ,
\end{equation}
\begin{equation}
\label{eq:finalB2}
\Gamma(B^{-}) - \Gamma(B^{0}) = \frac{G_{F}^{2} m_{b}^{2}}{4 \pi}
|V_{cb}|^{2} |V_{ud}|^{2} [A_{negint} - A_{exch} + {\cal
O}(1/m_{b})] \, .
\end{equation}
As $m_{b} \gg m_{c}$, it is clear that the approximation made above
should work much better for beauty particles than 
for charmed particles. Still, we assume that the approximations
made are justified in both $c$ and $b$ decays and that the corrections of
order ${\cal O}(1/m_{c,b})$ are not large. In this case, we obtain
the final relation for mesons:
\begin{equation}
\label{eq:meson}
r^{BD} \equiv \frac{\Gamma(B^{-}) - \Gamma(B^{0})}
{\Gamma(D^{+}) - \Gamma(D^{0})} = \frac{m_{b}^{2}}{m_{c}^{2}}
\frac{|V_{cb}|^{2}}{|V_{cs}|^{2}} + {\cal O}(1/m_{c},
1/m_{b}) \, .
\end{equation}

Let us consider relation (\ref{eq:meson}) in more
detail. By forming the lifetime differences (\ref{eq:diffD}) and 
(\ref{eq:diffB}) we have not only eliminated the effects of the leading 
operators in the mesonic decay rates, but applied  the {\em HQS}
 first at the subleading level of ${\cal O}(1/m_{c,b}^3)$. 
This has in turn enabled us to eliminate the dependence on 
four-quark matrix elements in relation (\ref{eq:meson}). 
Also, the sensitivity to the choice of 
heavy quark masses is now significantly reduced since in expression 
(\ref{eq:meson}) we have only the second power of masses.
 The procedure, however, has its limits.
It enables us to reduce the matrix elements, but
cannot be extended to {\em CKM} suppressed modes. Also, mass
corrections due to massive particles in the final decay states spoil
the procedure and, therefore, have been left out. 
Nevertheless, both of these corrections are under
good theoretical control.
Moreover, the contributions from mass corrections and suppressed modes
are smaller 
by more than one order of magnitude than the leading
contribution and cannot significantly change relation
(\ref{eq:meson}).

Starting from expression (\ref{eq:meson}), we may check the standard
formalism of inclusive decays against experimental data, 
especially if the four-quark operator
contributions are sufficient to explain the lifetime differences of
heavy mesons. To perform such a check, we determine
the quantity $r^{BD}$ using the experimental values for lifetimes  
and theoretically
(using relation (\ref{eq:meson})) and then compare the values.
Taking the experimental values from \cite{PDG}, we obtain 
\begin{equation}
\label{eq:rBDexp}
r^{BD}_{exp} = 0.030 \pm 0.011 \, .
\end{equation}
The large error of $r^{BD}_{exp}$ comes from the fact that the experimental
errors of lifetime measurements of $B^{-}$ and $B^{0}$ are large 
compared with the difference between central values of the results for 
$B^{-}$ and $B^{0}$. Inspection of \cite{PDG} shows that the
experimental situation in the sector of $B$ mesons is far from being settled
and more precise measurements of $B$ meson lifetimes are needed to
reduce the uncertainty in the value of $r^{BD}_{exp}$.

The theoretical value $r^{BD}_{th}$ is calculated from expression 
(\ref{eq:meson}). The numerical values for heavy-quark masses are
taken to be $m_{c}(m_{c}) = 1.25 \pm 0.1 \, GeV$ \cite{Uraltsev} and 
$m_{b}(1 \, GeV)= 4.59 \pm 0.08 \, GeV$ \cite{Bigi}
for the reasons discussed in \cite{GMS}. The values for the
matrix elements of the {\em CKM} matrix are taken from \cite{PDG} to be
$|V_{cs}| = 1.04 \pm 0.16$ and $|V_{cb}|=0.0402 \pm 0.0019$.
Using these numerical values we obtain
\begin{equation}
\label{eq:rBDth}
r^{BD}_{th} = 0.020 \pm 0.007 \, .
\end{equation}
The relatively large error in (\ref{eq:rBDth}) comes predominantly from
the large experimental error of the $|V_{cs}|$ {\em CKM} matrix element.

Direct comparison of the numerical results (\ref{eq:rBDexp}) and 
(\ref{eq:rBDth}) indicates that these are consistent
within errors. Since the relative errors of both results are rather
large and experimental data are still fluid, it is possible that
these results will experience some change.
Nevertheless, 
we expect that the two results will remain comparable and
even closer to each other. 

These results confirm, in a model-independent way, that four-quark
operators can account for the greatest part of 
decay rate differences of heavy mesons, i.e., that the contributions
of other operators of dimension $6$ and
higher-dimensional operators are not so important. Of course, these
conclusions should be interpreted in the light of the approximations
made.

The procedure presented for heavy mesons
can be extended to heavy baryons. First, we consider the system
of singly heavy baryons. These baryons contain two light quarks,
which introduces two types of four-quark operator contributions for
each baryon. Again, we choose two pairs of singly heavy baryons,
each pair containing one charmed and one beauty baryon related by the same 
dominant type of four-quark contributions.  The first
pair contains $\Xi_{c}^{+} \, (cus)$ and $\Xi_{b}^{-} \, (bds)$,
 while the second pair contains $\Xi_{c}^{0} \, (cds)$ and
$\Xi_{b}^{0} \, (bus)$. The first pair exhibits the negative Pauli
interference and different nonleptonic and semileptonic contributions of 
four-quark operators containing $s$ quark fields. 
The second pair comprises the $W$-exchange effect 
and the same contributions of four-quark operators
containing $s$ quark fields as in the first pair of baryons. 
It is important to notice that
the four-quark operators that contain $s$ quark fields do not describe
the same type of effects in $c$ and $b$ decays in our pairs 
(and therefore are
not given by the contributions of the same type in the effective Lagrangian).
Nevertheless, we form decay-rate differences in such a manner that
the contributions of four-quark operators containing the $s$-quark
field cancel (assuming {\em isospin} symmetry). This cancellation is
important in the case of $b$ decays because in the contributions of
$s$-quark type four-quark operators significant mass
corrections appear because of two massive $c$ quarks in the final
state. 

In the singly heavy sector we, therefore, 
form the following differences:
\begin{equation}
\label{eq:diffXic}
\Gamma(\Xi_{c}^{+}) - \Gamma(\Xi_{c}^{0}) = \frac{G_{F}^{2} m_{c}^{2}}{4 \pi}
|V_{cs}|^{2} |V_{ud}|^{2} [ \langle \Xi_{c}^{+} |
 P^{cu} | \Xi_{c}^{+} \rangle -
\langle \Xi_{c}^{0} | P^{cd} |\Xi_{c}^{0} \rangle ]\, ,
\end{equation}
%
\begin{equation}
\label{eq:diffXib}
\Gamma(\Xi_{b}^{-}) - \Gamma(\Xi_{b}^{0}) = \frac{G_{F}^{2} m_{b}^{2}}{4 \pi}
|V_{cb}|^{2} |V_{ud}|^{2} [ \langle \Xi_{b}^{-} | 
P^{bd}| \Xi_{b}^{-} \rangle -
\langle \Xi_{b}^{0} | P^{bu} | \Xi_{b}^{0} \rangle ] \, .
\end{equation}

Following the procedure displayed in relations
(\ref{eq:HQSfornegintD}) to (\ref{eq:meson}) and
using the {\em isospin} symmetry and {\em HQS} again, we obtain the
expressions (note that the four-quark operators which contribute
to the negative Pauli interference in mesons contribute to $W$
exchange in baryons and vice versa \cite{cmesons,cbaryons}):
%
%
%
%
\begin{equation}
\label{eq:finalXic}
\Gamma(\Xi_{c}^{+}) - \Gamma(\Xi_{c}^{0}) = \frac{G_{F}^{2} m_{c}^{2}}{4 \pi}
|V_{cs}|^{2} |V_{ud}|^{2} [B_{negint} - B_{exch} + {\cal
O}(1/m_{c})] \, ,
\end{equation}
and
\begin{equation}
\label{eq:finalXib}
\Gamma(\Xi_{b}^{-}) - \Gamma(\Xi_{b}^{0}) = \frac{G_{F}^{2} m_{b}^{2}}{4 \pi}
|V_{cb}|^{2} |V_{ud}|^{2} [B_{negint} - B_{exch} + {\cal
O}(1/m_{b})] \, ,
\end{equation}
where the matrix elements of the four-quark operators between 
singly heavy baryons $B_{Qqq'}$ have the form $B_{negint} = \langle B_{Qqq'} |
\sum_{i=1}^{4} c_{i, negint}^{Qq} (\mu) O_{i, negint}^{Qq} (\mu)
| B_{Qqq'} \rangle$ and $B_{exch} = \langle B_{Qqq'} |
\sum_{i=1}^{2} c_{i, exch}^{Qq} (\mu) O_{i, exch}^{Qq} (\mu)
| B_{Qqq'} \rangle$. 

The final relation for singly heavy baryons looks like 
\begin{equation}
\label{eq:rbc}
r^{bc} \equiv \frac{\Gamma(\Xi_{b}^{-}) - \Gamma(\Xi_{b}^{0})}
{\Gamma(\Xi_{c}^{+}) - \Gamma(\Xi_{c}^{0})} =
\frac{m_{b}^{2}}{m_{c}^{2}} \frac{|V_{cb}|^{2}}{|V_{cs}|^{2}}
[1 + {\cal O}(1/m_{c}, 1/m_{b})] \, .
\end{equation}

This relation can be used to test the present model-dependent 
calculation of heavy-baryon lifetimes. Lifetime splittings 
both in the charmed and in the beauty sector have been investigated 
in several papers \cite{Neu,BlokShif,GM,Voloshin,GMS,Vol}. 
We have recalculated our predictions 
from \cite{GM} for charmed baryon lifetimes and from \cite{GMS} 
for 
beauty baryon lifetimes with the numerical parameters preferred in this 
paper and have used the recalculated predictions 
to test the relation (\ref{eq:rbc}),
which we express in units $\frac{m_{b}^{2}}{m_{c}^{2}} 
\frac{|V_{cb}|^{2}}{|V_{cs}|^{2}}$
thus reducing the dependence on heavy-quark masses.
In the case when the same approximations (neglecting 
 mass corrections and Cabibbo-suppressed modes) are made, 
the above relation differs from unity by $-12\%$. This number shows 
the deviation of the model-dependent calculation (where the contributions 
of four-quark operators are explicitly evaluated) 
 from the model-independent prediction given only by the ratios of 
heavy-quark 
masses and Cabibbo matrix elements. 
The complete 
calculation with the mass corrections and Cabibbo-suppressed modes 
gives $0.79$ for the above ratio, which indicates the order of 
neglected corrections to be less than $10\%$. 

There are several important implications of relation (\ref{eq:rbc}). 
We can obtain a model-independent prediction for the still 
unmeasured lifetime difference between singly beauty baryons in  
the isospin doublet. Using the same numerical values for the 
parameters entering the right-hand side as before, and 
taking the experimentally 
measured lifetimes of singly-charmed baryons $\Xi_{c}^{+}$ and 
$\Xi_{c}^{0}$ from \cite{PDG}, we obtain
\begin{equation}
\label{eq:diffSBB}
\Gamma(\Xi_{b}^{-}) - \Gamma(\Xi_{b}^{0}) =-(0.14 \pm 0.06) \quad 
{\rm ps^{-1}} \; . 
\end{equation}

This lifetime difference can be compared with the model-independent 
prediction obtained by using the {\em HQS} and $SU(3)$ symmetry 
in \cite{Voloshin},  
$\Gamma(\Xi_{b}^{-}) - \Gamma(\Xi_{b}^{0}) =-(0.11 \pm 0.03)
{\rm ps^{-1}}$,
and some moderate model-dependent prediction from \cite{GMS}, 
$\Gamma(\Xi_{b}^{-}) - \Gamma(\Xi_{b}^{0}) =-0.094 \, {\rm
ps^{-1}}$.
It is also worth mentioning that owing to the neglected mass 
corrections in final decay states of beauty baryons in the derivation of the 
ratios (\ref{eq:rbc}) and (\ref{eq:diffSBB}), there is no splitting 
between the nonleptonic rates of $\Xi_{b}^{0}$ and $\Lambda_{b}$.
Since the Cabibbo-suppressed modes have also been 
neglected , the total decay rates of 
$\Xi_{b}^{0}$ and $\Lambda_{b}$ appear to be equal at this level. 
Therefore, the predictions from 
(\ref{eq:rbc}) are also valid for the 
$\Gamma(\Xi_{b}^{-}) - \Gamma(\Lambda_{b})$ difference. 
We can see that the results for the predicted splitting in the 
isospin doublet of beauty baryons obtained using different methods are 
all consistent. 

It is interesting to examine the explicit expression 
taken from \cite{Voloshin}, eq.(23), which can be rewritten as
\begin{equation}
\label{eq:rbcV}
\Gamma(\Xi_{b}^{-}) - \Gamma(\Xi_{b}^{0}) = 
\frac{m_{b}^{2}}{m_{c}^{2}} \frac{|V_{cb}|^{2}}{|V_{cs}|^{2}}
[0.91 \Gamma(\Xi_{c}^{+}) - 0.85 \Gamma(\Xi_{c}^{0}) - 
0.06 \Gamma(\Lambda_{c})] \, , 
\end{equation}
and which is obtained by applying $SU(3)$ symmetry, and with Cabibbo 
subleading effects included in charmed baryon decays. 
The appearance of the total decay rate for $\Lambda_{c}$ is the direct 
consequence of the $SU(3)$ relations used. Otherwise, we can note a
very similar structure to our relation (\ref{eq:rbc}) obtained within 
the $SU(2)$ approximation.  The first two 
coefficients in front of the rates in 
(\ref{eq:rbcV}) differ from one by less than $15\%$, where about 
$5-10\%$ difference comes from the Cabibbo-suppressed modes included, 
and the rest comes presumably from the difference between the $SU(3)$ and $SU(2)$ 
approximations applied.  

In the last case, we apply our procedure to the system of
doubly heavy baryons. Again, we form pairs of one doubly charmed and
one doubly beauty baryon. The first pair includes $\Xi_{cc}^{++} \,
(ccu)$ and $\Xi_{bb}^{-} \, (bbd)$. The decays of both baryons in
this pair include the effect of negative Pauli interference. The
second pair is formed from  $\Xi_{cc}^{+} \, (ccd)$ and
$\Xi_{bb}^{0} \, (bbu)$. Both of these baryons exhibit the effect of 
$W$ exchange. There are no Cabibbo-leading semileptonic 
four-quark contributions in doubly heavy baryons. 
Appropriate decay-rate differences are given by the
following expressions:
\begin{equation}
\label{eq:diffXicc}
\Gamma(\Xi_{cc}^{++}) - \Gamma(\Xi_{cc}^{+}) = \frac{G_{F}^{2} m_{c}^{2}}{4 \pi}
|V_{cs}|^{2} |V_{ud}|^{2} [ \langle \Xi_{cc}^{++} | 
P^{cu}| \Xi_{cc}^{++} \rangle -
\langle \Xi_{cc}^{+} | P^{cd} | \Xi_{cc}^{+} \rangle ]\, ,
\end{equation}
\begin{equation}
\label{eq:diffXibb}
\Gamma(\Xi_{bb}^{-}) - \Gamma(\Xi_{bb}^{0}) = \frac{G_{F}^{2} m_{b}^{2}}{4 \pi}
|V_{cb}|^{2} |V_{ud}|^{2} [ \langle \Xi_{bb}^{-} | 
 P^{bd}| \Xi_{bb}^{-} \rangle -
\langle \Xi_{bb}^{0} |  P^{bu}
| \Xi_{bb}^{0} \rangle ] \, .
\end{equation}

Using {\em isospin symmetry} and {\em HQS}, we obtain,
per analogiam with above derivations, relations for
the decay-rate differences of doubly heavy baryons
%
%
%
%
which lead to the final relation for doubly heavy baryons
\begin{equation}
\label{eq:rbbcc}
r^{bbcc} \equiv \frac{\Gamma(\Xi_{bb}^{-}) - \Gamma(\Xi_{bb}^{0})}
{\Gamma(\Xi_{cc}^{++}) - \Gamma(\Xi_{cc}^{+})} =
\frac{m_{b}^{2}}{m_{c}^{2}} \frac{|V_{cb}|^{2}}{|V_{cs}|^{2}}
[1 + {\cal O}(1/m_{c}, 1/m_{b})] \, .
\end{equation}
Again, we have achieved the reduction of the matrix elements 
of four-quark operators 
between doubly heavy baryons in the heavy quark limit.

Relation (\ref{eq:rbbcc}) enables us to estimate the difference in
the decay rates of doubly-beauty baryons using recent results 
\cite{GMScc} for doubly-charmed baryon lifetimes. Using the calculated
values
$\Gamma(\Xi_{cc}^{++}) = 0.952 \, ps^{-1}$ and 
$\Gamma(\Xi_{cc}^{+}) = 5 \, ps^{-1}$ with the values of the parameters
$m_{c} = 1.35 \, GeV$, $m_{b}=4.7 \, GeV$, $|V_{cb}| = 0.04$
and $|V_{cs}| = 1.04$ gives
\begin{equation}
\label{eq:diffnumXibb}
\Gamma(\Xi_{bb}^{-}) - \Gamma(\Xi_{bb}^{0}) = - 0.073 \, 
{\rm ps^{-1}} \, .
\end{equation}

So far we have applied our reduction procedure separately to heavy
mesons, singly-heavy baryons, and doubly-heavy baryons. Still, by
inspection of relations (\ref{eq:meson}), (\ref{eq:rbc}), and 
(\ref{eq:rbbcc}) we see that $r^{BD}$, $r^{bc}$, and $r^{bbcc}$ are
the same up to ${\cal O}(1/m_{c},1/m_{b})$, i.e.,
\begin{equation}
\label{eq:connect}
\frac{\Gamma(B^{-}) - \Gamma(B^{0})}{\Gamma(D^{+}) - \Gamma(D^{0})}
= \frac{\Gamma(\Xi_{b}^{-}) - \Gamma(\Xi_{b}^{0})}
{\Gamma(\Xi_{c}^{+}) - \Gamma(\Xi_{c}^{0})} =
\frac{\Gamma(\Xi_{bb}^{-}) - \Gamma(\Xi_{bb}^{0})}
{\Gamma(\Xi_{cc}^{++}) - \Gamma(\Xi_{cc}^{+})} =
\frac{m_{b}^{2}}{m_{c}^{2}} \frac{|V_{cb}|^{2}}{|V_{cs}|^{2}} \, .
\end{equation}

Thus, we obtain a relation in 
(\ref{eq:connect}) which clearly indicates certain universal
behavior in the decays of all heavy hadrons. The existence of some
universality could have been anticipated from the fact that the
same expression (\ref{eq:master}) describes the decays of all heavy
baryons. Still, that universality attains  
its concrete, model-independent form
in the relation (\ref{eq:connect}). This
relation connects all sectors of heavy hadrons that are usually
treated separately: mesonic and baryonic, charmed and beauty.
Also, this relation brings some order in the otherwise rather
intricate pattern of heavy-hadron lifetimes. 
An advantage of this relation is that by
knowing some decay rates, one can calculate or give constraints on
some other decay rates. Also, knowing decay rates from
experiment, one can test the findings of the theory in a 
model-independent fashion.
These results will help to
establish the limitations of the present standard method of
calculating inclusive processes and to test its
underlying assumptions.

{\bf Acknowledgements}

This work was supported by the Ministry of
Science and Technology of the Republic of Croatia under the
contract No. 00980102.

\end{document}